\documentclass{article}

\usepackage{arxiv}

\usepackage[utf8]{inputenc} 
\usepackage[T1]{fontenc}    
\usepackage{hyperref}       
\usepackage{url}            
\usepackage{booktabs}       
\usepackage{amsfonts}       
\usepackage{nicefrac}       
\usepackage{microtype}      
\usepackage{lipsum}
\usepackage{graphicx}
\usepackage{placeins}

\title{Inverse Design of Photonic Metasurface Gratings for Beam Collimation in Opto-fluidic Sensing}

\author{
  Robin Singh\thanks{robinme@mit.edu}\\
  Department of Mechanical Engineering\\
  Massachusetts Institute of Technology\\
  Cambridge, MA, 02139, USA \\
   \And
 Yuqi Nie \\
  Department of Materials Science and Engineering\\
  Massachusetts Institute of Technology\\
  Cambridge, MA, 02139, USA \\
   \And
 Anuradha Murthy Agarwal \\
  Department of Materials Science and Engineering\\
  Massachusetts Institute of Technology\\
  Cambridge, MA, 02139, USA \\
   \And
 Brian W Anthony\thanks{banthony@mit.edu} \\
  Department of Mechanical Engineering, MIT.nano\\
  Institute for Medical Engineering and Science\\
  Massachusetts Institute of Technology\\
  Cambridge, MA, 02139, USA \\
}

\begin{document}
\maketitle

\begin{abstract}
Metasurfaces provide the disruptive technology enabling miniaturization of complex cascades of optical elements on a plane. We leverage the benefits of such a surface to develop a planar integrated photonic beam collimator for on-chip optofluidic sensing applications. While most of the current work focuses on miniaturizing the optical “\emph{detection}” hardware, little attention is given to develop on-chip hardware for optical “\emph{excitation}”. In this manuscript, we propose a flat metasurface for beam collimation in optofluidic applications. We implement an inverse design approach to optimize the metasurface using gradient descent method and experimentally compare its characteristics with conventional binary grating-based photonic beam diffractors. The proposed metasurface can enhance the illumination efficiency almost two times in on-chip applications such as fluorescence imaging, Raman and IR spectroscopy and can enable multiplexing of light sources for high throughput biosensing.
\end{abstract}

\section{Introduction}

Silicon photonics promises highly effective, low-cost, and scalable solutions to miniaturize electronic and photonic systems [1-3], enabling on-chip spectroscopic sensing and imaging techniques in the fields of medicine and biology [4-6], for examining the behavior of biomolecular species and living cells. Among these techniques, the most common are fluorescence imaging/microscopy, infra-red (IR) spectroscopy, and Raman spectroscopy [7-13]. Fluorescence imaging/microscopy is a powerful tool for biomedical research because it provides very high sensitivity and specificity for cellular activity detection, making it the gold standard. Over the past decade [14-15], with the advent of semiconductor image sensing technology, researchers have demonstrated on-chip contact-based fluorescence detection techniques with high throughput and scalability [12]. Takeshara et al. implemented contact fluorescence microscopy in microfluidic chips [14]. Pang et al. developed various optofluidic devices that can perform fluorescence detection, imaging and sensing [8]. 

Similarly, IR and Raman spectroscopic techniques based on phononic vibrational states of molecules are  powerful methods to analyze their chemistry.  The fact that these vibrational spectroscopic techniques are label-free methods, has evoked a renewed interest among researchers to miniaturize them to develop high throughput and high resolution \emph{lab-on-chip} sensors. Perichetti et al. developed a multifunctional platform for Raman and fluorescence spectroscopic analysis [13]. Chen et al. investigated on-chip methods for Surface Raman spectroscopy integrated with thin layer chromatography [16]. Further, IMEC recently demonstrated SiN waveguide-based on-chip Raman spectroscopy that promises high throughput and better spectral resolution [17]. 

The research to date in terms of miniaturizing these techniques tends to focus more on developing hardware for on-chip “detection”. However, far too little attention is paid to on-chip hardware for “excitation”.  Most of the existing work uses benchtop laser beams focused on the chip for excitation [8, 14]. A major disadvantage of this method is that photonic excitation is not spatially confined and results in high background noise, leading to poor signal-to-noise (SNR) ratio and sensitivity. 

If we can develop a platform that can miniaturize the excitation and detection hardware on a small chip, it would offer a multitude of advantages. First, scaling up of the imaging field of view can be achieved by multiplexing a large number of excitation sources. Second, it provides compact hardware geometry to implement the setup at a low-cost with high volume production. Third, it offers opportunities to develop automated methods for biological analysis such as cell screening and probing. 

Our planar photonic waveguide-based metasurface can enable miniaturized excitation sources which efficiently guide and manipulate photons on the chip. Researchers have previously demonstrated waveguides which guide and diffract light out of the plane using conventional grating surfaces. For example, Kerman et al used a conventional focused grating coupler design to excite samples in a microfluidic chip [18]. Such designs typically enhance fiber coupling in photonic integrated circuits [19-21]. However, these gratings employ a high degree of symmetry in their structure and so fail to a provide uniform collimated out of-the-plane excitation of the liquid sample flowing in an integrated microfluidic sensor channel. An artificial diffracting (meta) surface that can collimate the beam out of the plane uniformly over a wide area would greatly benefit on-chip microfluidic sensing [19,22-25], but meta-gratings today only focus on improving fiber to chip coupling.

In the present manuscript, we use an inverse design approach to obtain a metasurface based photonic grating structure that provides a wide collimated beam for on-chip optical excitation and detection [26-27] for bio-sensing applications. We present FDTD modeling, optimization, fabrication and demonstration of a meta collimator, and compare it with conventional grating coupling designs.

\begin{figure}[htbp]
\centering
\includegraphics[scale=0.5]{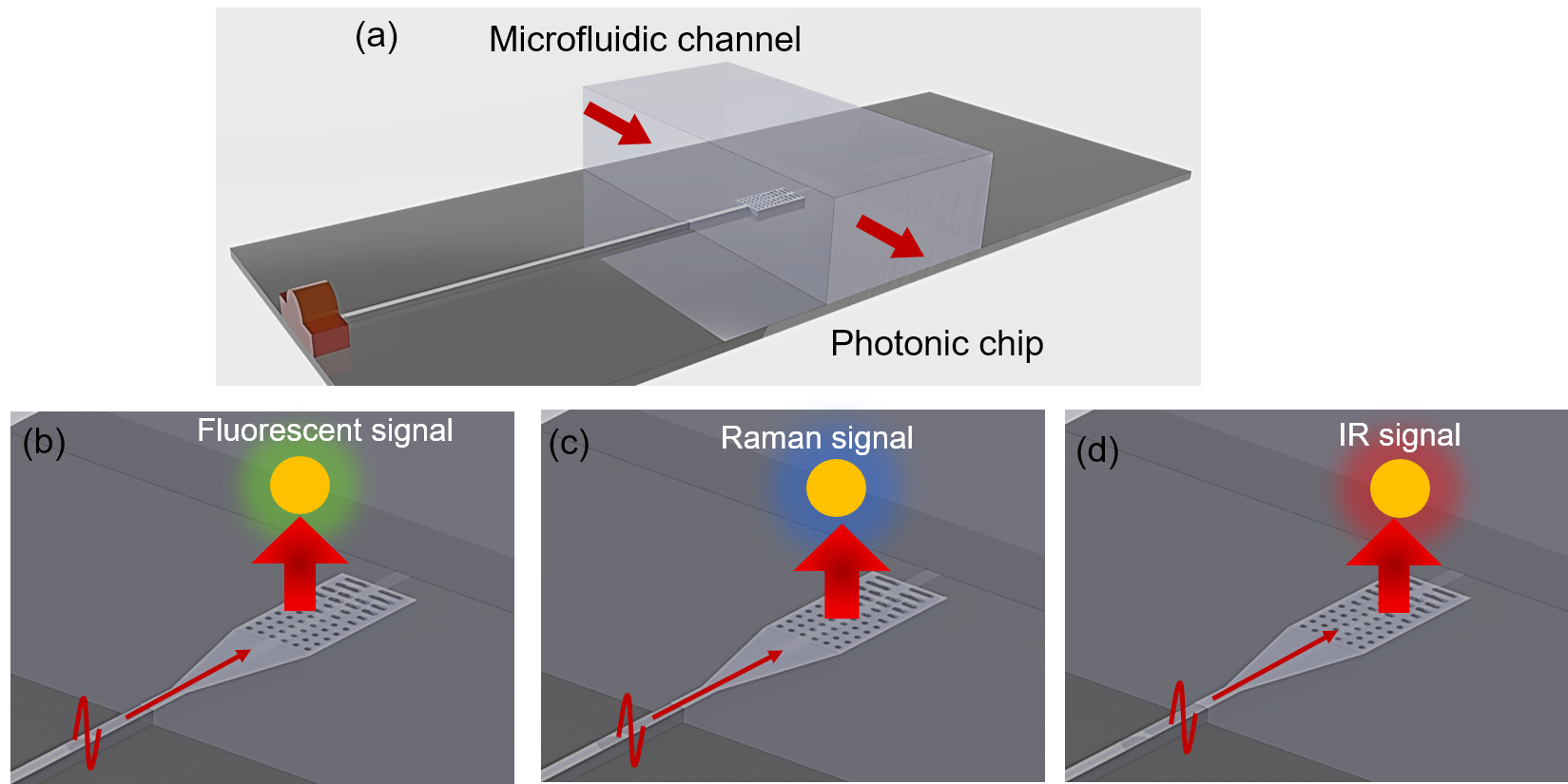}
\caption{\textit{Opto-fluidic Lab on the chip sensing system. (a) Photonic waveguide technology integrated with a microfluidic chip to perform high throughput detection and sensing. (b) Planar photonic waveguide based light excitation and detection for implanting fluorescence spectroscopy. The excitation of the sample (shown in red) generates a fluorescent signal (shown in green) detected by adjoining detectors (c) Planar waveguide-based light excitation for on-chip Raman spectroscopy. The Raman signals are represented in blue. (d) Planar waveguide-based light excitation for on-chip IR spectroscopy where excitation and detection signals are shown in the red.} }
\label{fig1}
\end{figure}

\section{Meta Collimator Design, Modeling and Fabrication}
\label{sec:headings}

Conventional waveguide-based gratings have multiple identical diffracting grooves along the direction of light propagation that diffracts the photonic beam out-of-the-plane at a specified angle (Fig. 2c). However, the emission power of the light decreases exponentially as it progresses along the waveguide. To achieve uniform emission along the waveguide, we need to control the amount of energy diffracted from each individual diffracting groove, which is accomplished by engineering a unique distribution of light-scatterers along the direction of light propagation. We define rectangular light scattering grooves in the grating structure by their duty cycle (dc), row period ($\Lambda_v$) along the transverse direction, and line period ($\Lambda_h$), as shown in Fig. 2b.   

As a demonstration, we maintain the width, $w$, and length, $l$ of the grating structure as 10 $\mu$ m and 20 $\mu$ m respectively (Fig. 2a).  We optimize the distribution of scatterers to obtain uniform emission from the grating structure. The center wavelength of the emission spectrum is set to be in the C-band (1550 nm). Further, we ensure that the device can operate over a wide ($\sim$70 nm) range of wavelengths while optimizing its structure for uniformity. 

In our methodology, the optimization of the meta-grating is divided into two stages. The first optimization stage initializes the duty cycle of an individual row in the grating structure and then performs an iterative gradient descent inverse optimization to collimate the output beam. In general, gradient descent methodologies have many advantages. It allows us to optimize the photonic structure with relatively large degrees of freedom as compared to other gradient-free optimization schemes. Further, it requires fewer simulation steps and does not rely on parametric sweeps or random mathematical perturbations to find the optimum values.

\begin{figure}[htbp]
\centering
\includegraphics[scale=0.3]{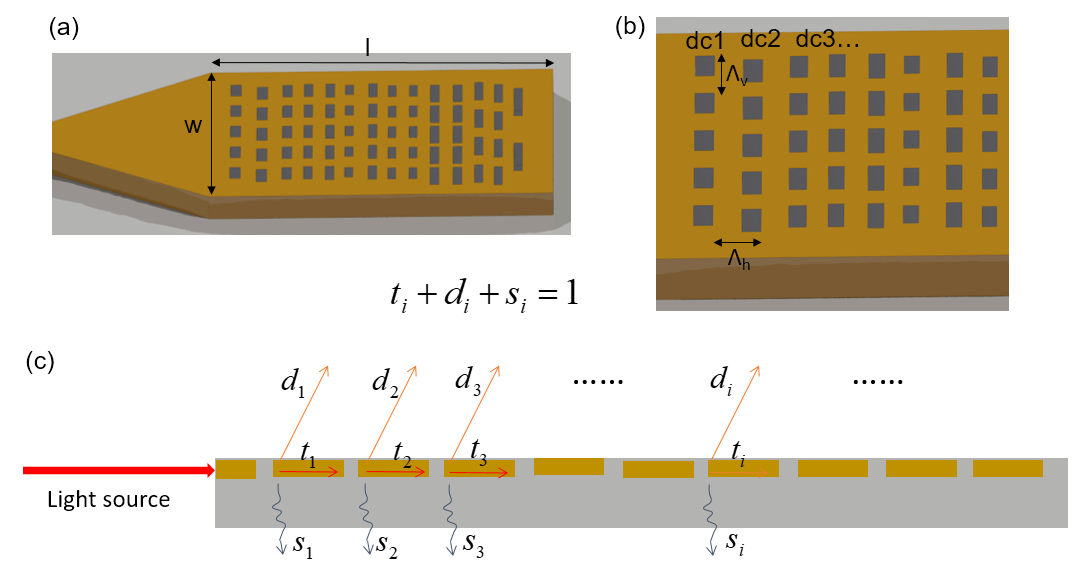}
\caption{\textit{(a) Meta grating structure with rectangular scatterers in each grating row. (b) The rectangular scatterers are dispersed within a row and their characteristics are determined by the duty cycle (dc) and their period ($\Lambda_v$). Each row has a different characteristic of rectangular scatterers that determine how much they diffract the light off the plane to illuminate the sample. (c) Light propagation in the metasurface is easily understood through effective mirror model where the light gets diffracted (d), absorbed (s) and transmitted (t) through the individual row.}}
\label{fig2}
\end{figure}

\subsection{Parameter Initialization}
As the gradient descent method is highly sensitive to initial conditions, it is customary to initialize the parameters with relevant values. To come up with the initial values, we use an \emph{effective mirror model} for the meta grating structure. The grating grooves can be approximated by the cascaded mirror model to understand the light propagation through the structure [28]. Fig. 2c shows the meta grating structure with its cascade mirror model. Transmission, diffraction, and scattering coefficients of the diffracting groove are assumed as $t$, $d$, and $s$ respectively. Diffraction intensity output from the first diffracting groove is proportional to $d$, and is given by, 

\begin{equation}
I_1=d_1
\label{eq1}
\end{equation}

Using the cascaded mirror model, we can write intensity $I_2$ and $I_n$ in general are given by, 

\begin{equation}
I_2=t_1 d_2
\label{eq2}
\end{equation}

\begin{equation}
I_n=t_1 t_2 ... t_{n-1} d_n
\label{eq3}
\end{equation}

The collimated beam requires uniform emission from an individual groove. Mathematically, the condition is given by,

\begin{equation}
I_1=I_2= ... I_n
\label{eq4}
\end{equation}

Substituting the equations (\ref{eq1}) , (\ref{eq2}), (\ref{eq3}) in the condition  (\ref{eq4}) results in

\begin{equation}
d_1=t_1 d_2 =t_1 t_2 d_3 =...=t_1 t_2 ... t_{n-1} d_n
\label{eq5}
\end{equation}

Assuming $f_i$, $d_i$ the functions of the duty cycle, x and width w, of the groove. For constant width, we sweep the duty cycle and obtain the functions, $f_i (x)$ and $d_i (x)$. Fig. 3a shows both the functions plotted for different values of the duty cycle, x. 

Using the functions’ plot , we initialize the duty cycle for individual grating groove to satisfy the condition given by, 

\begin{equation}
f_{i-1}=\frac{d_{i-1}}{t_{i-1}}=d_i
\label{eq6}
\end{equation}

\begin{figure}[htbp]
\centering
\includegraphics[scale=0.4]{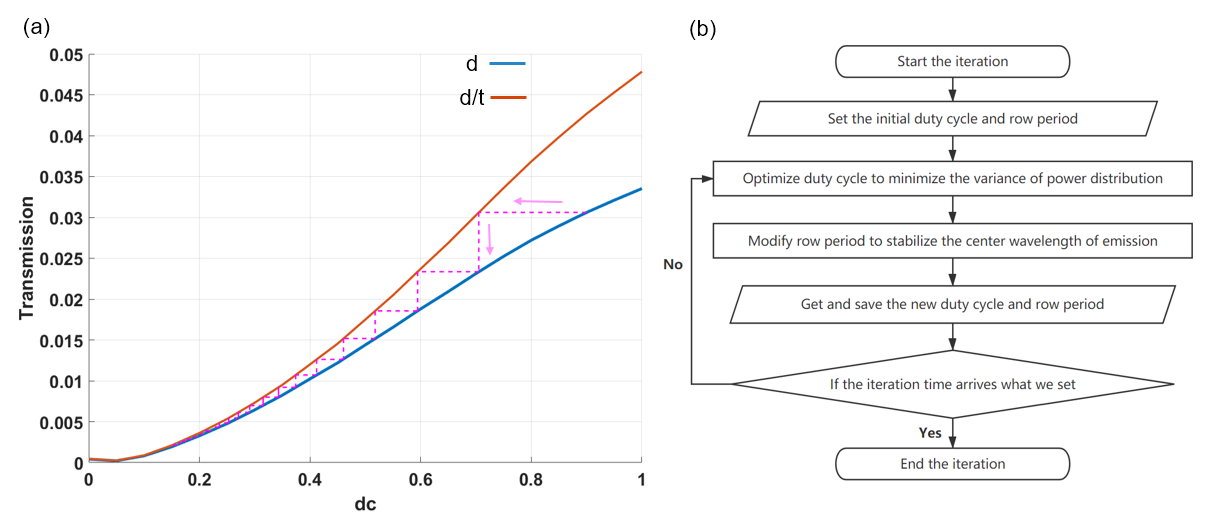}
\caption{\textit{This is the control flow chart of the iteration algorithm. We first set the initial duty cycle and row period in the grating structure, then use gradient descent method to tune the duty cycle and optimize the variance of power distribution. This may cause a shift in central wavelength, thus we modify the groove period to stabilize it. So we get the new duty cycle and groove period in every loop and iterate it till a uniform power and required waveband collimator is achieved.}}
\label{fig3}
\end{figure}

\subsection{Iteration Method}
After the initialization process as explained above, we optimize the duty cycle of the individual grating groove in the meta-surface and obtain a collimated beam output. Lumerical FDTD software is used to perform the optimization which is based on the fact that the diffracted power from an individual groove increases with an increase in its duty cycle. We calculate the spatial distribution of diffracted power and its variance, then tune the duty cycle of each groove to minimize the variance. Fig. 3b shows the control flow chart of our iterative algorithm. With every iteration, the uniformity in the power distribution of the beam collimation increases. When the duty cycle and the grating period change, the effective index of the structure changes, as a result of which the center wavelength of grating emission also shifts. This may result in poor convergence of the duty cycle in the iterative procedure. Therefore, the central emission wavelength needs to be stabilized in the iterations by modifying the grating period. We add an additional step in our algorithm to tune the groove period ($\Lambda_h$) and stabilize the emission wavelength. Results obtained from the optimization process are shown in Fig 4. It shows the top and side view of the output beam from the optimized metasurface and compares it with a conventional binary grating structure. 

After optimization, we get a collimated beam with up to ~50nm of operating bandwidth. Besides, it should be noted that some spatial randomness is incorporated in each groove to avoid the lattice diffraction patterns in the profile. More details on the iterative procedure followed in the optimization can be found in the Appendix of the manuscript.

\begin{figure}[htbp]
\centering
\includegraphics[scale=0.35]{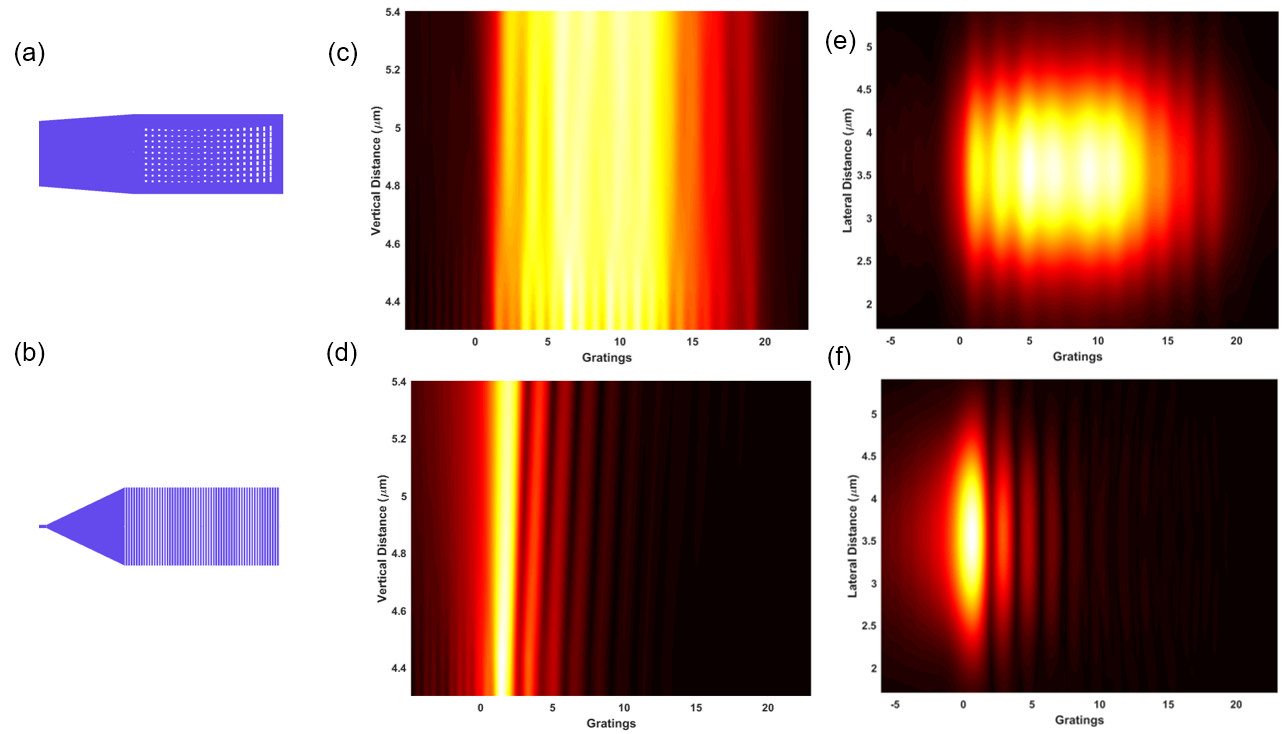}
\caption{\textit{FDTD simulation of optimized meta grating and the conventional binary gratings. (c, d) Side view of the diffracted beam profile that illuminates the analyte flowing over the plane for the optimized meta and conventional gratings respectively. (e, f) Top view of the beam spot obtained from the meta and conventional grating structures respectively.} }
\label{fig4}
\end{figure}

With the optimized design parameters, we fabricate the structures using the fabrication steps shown in Fig. 5. A low-pressure chemical vapor deposition (LPCVD) system is used to deposit 400 nm thick silicon nitride ($Si_3 N_4$) layer on a 6-inch silicon dioxide wafer (3-micron oxide on Si substrate). These thermal oxide wafers are procured from Wafer Pro LLC, CA. The gratings and waveguides are then designed and patterned on silicon nitride on-insulator substrate via e-beam lithography and followed by reactive ion etching to define the geometry of the grating structures. Fluorine chemistry with a gas mixture of CHF$_3$ and CF$_4$ is used in the dry etching step. The leftover resist is stripped off using oxygen cleaning and acetone rinsing. 

\begin{figure}[htbp]
\centering
\includegraphics[scale=0.4]{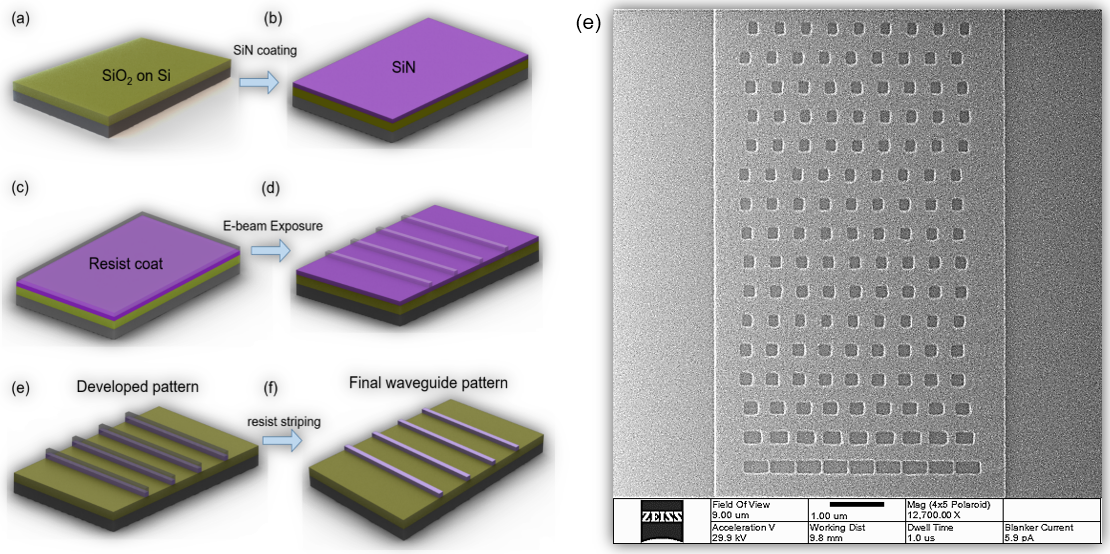}
\caption{\textit{Process fFlow steps involved in the fabrication of the metasurface. (b) He ion microscopy of the patterned structure on topin the top of sSilicon nitride film.}}
\label{fig5}
\end{figure}

\section{Design Performance and Characterization}
\label{sec:others}

We benchmark the fabricated photonic meta-surface against a conventional binary grating-based diffractor. Fig. 6 shows the experimental analysis of the beam profile for the micro-structures. Please note that we display a 1.5x zoomed version of the beam profile from the conventional gratings for better visualization against the 1x version of the beam profile from the meta structure.  During the experimental characterization, the CinCam InGasAS SWI camera (from Axiom Optics) is mounted upside down to analyze the beam. Fig 5b shows the microscopic view of the light propagating in the photonic waveguides.  Fig. 6c and e compare the side view of the beam pattern, whereas the respective top views are shown in Figs. 6d and 6f. We find that the power delivered through the meta-surface is almost 1.4 times that of the conventional grating surface. The SNR of the illumination increased from 15 dB to 23 dB.  The beam spot also improves from ROI of 0.06 mm$^2$ to 0.1 mm$^2$. Further, the homogeneity of the beam increases almost 2 times with the meta-surface enabling greater light illumination efficiency of the on-chip device.

\begin{figure}[htbp]
\centering
\includegraphics[scale=0.45]{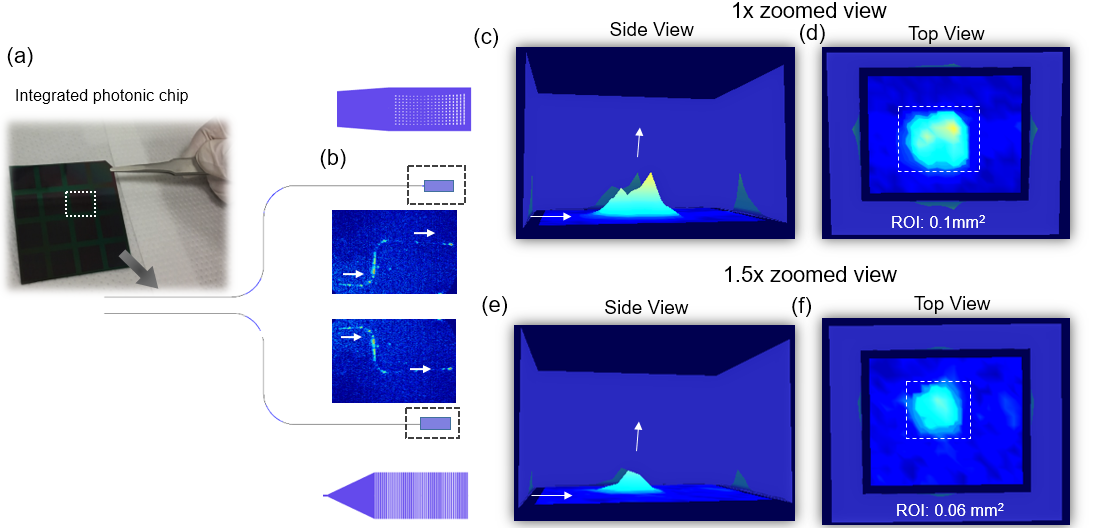}
\caption{\textit{(a, b) Experimental comparison of metasurface and conventional binary grating-based diffractor. a) The fabricated sample to test. (b) Microscopic top view of the light propagating the photonic waveguides.  (c, d, e, f) Side and top view of the beam profiles from the metasurface and the binary grating structure.}}
\label{fig6}
\end{figure}

\section{Conclusion}
In this manuscript, we demonstrate a meta-surface based photonic beam collimator for on-chip optofluidic probing. With renewed interest in developing chip-scale spectroscopic methods for sensing, the beam collimator would be an attractive option for miniaturized excitation sources.  We adopt an inverse design strategy to optimize the meta structure. The optimization was based on a gradient descent method where the parameters were initialized using an effective mirror model. The designed meta-surface is successfully fabricated and experimentally characterized. We compare its beam profiling performance with a conventional binary grating structure. Such an excitation source when integrated with a planar waveguide-based photonic detector has the potential to miniaturize on-chip spectroscopy. 

\section{Appendix}
We explain the optimization process in greater detail here. As stated earlier in the manuscript, the optimization parameters are the duty cycle and the grating period. The duty cycle optimization is explained in the flow chart given in Fig. 7. The flow chart for the row period optimization is shown in Fig. 8.

\FloatBarrier
\begin{figure}[htbp]
\centering
\includegraphics[scale=0.6]{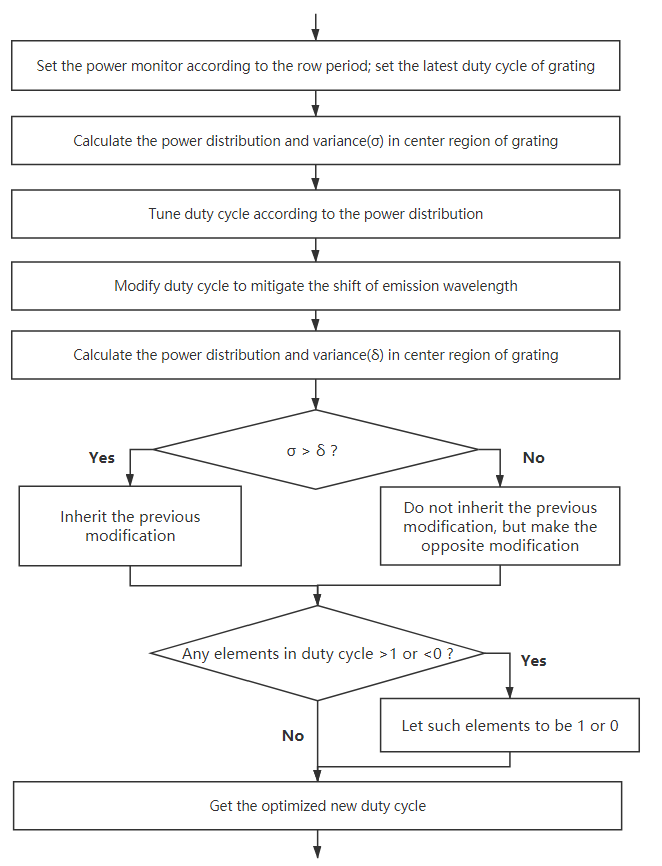}
\caption{Detailed control flow chart of “Optimize dc to minimize the variance of power distribution” in Fig.3b.}
\label{fig7}
\end{figure}
\FloatBarrier

\FloatBarrier
\begin{figure}[htbp]
\centering
\includegraphics[scale=0.4]{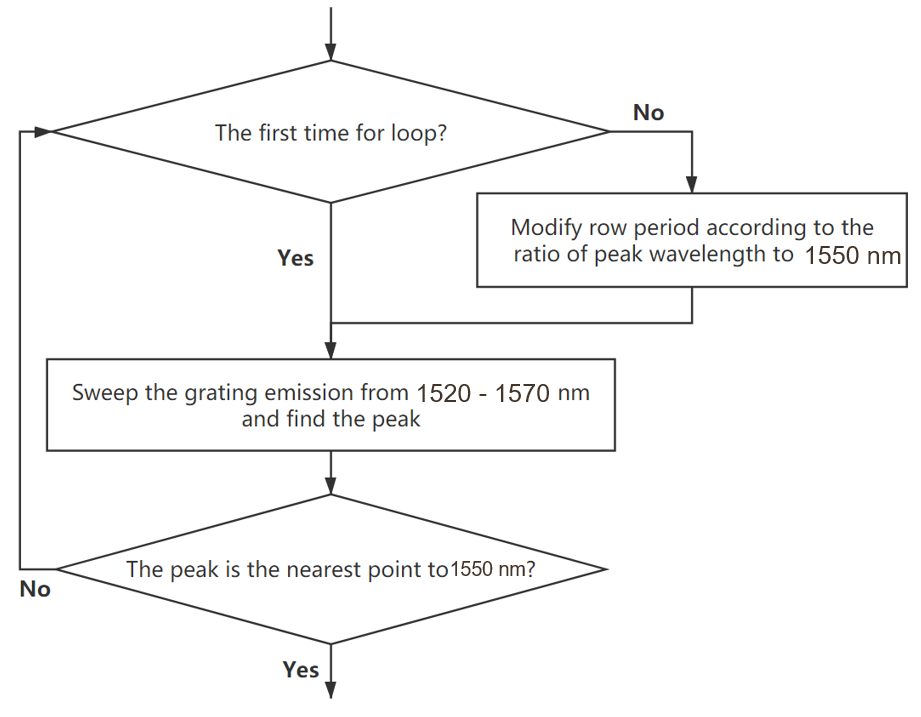}
\caption{Detailed control flow chart of the “Modify groove period to stabilize the center wavelength of emission” in Fig.3b.}
\label{fig8}
\end{figure}
\FloatBarrier

\bibliographystyle{unsrt}  


\end{document}